# Reconstruction of an effective magnon mean free path distribution from spin Seebeck measurements in thin films (DOI: 10.1088/1367-2630/aa5163)

E. Chavez-Angel[*,1], R. A. Zarate[2], S. Fuentes[3,4], E. J. Guo[1,5], M. Kläui[1,6] and G. Jakob[1,6]

[1] Institute of Physics, University of Mainz, Staudinger Weg 7, 55128 Mainz, Germany.

[2] Depto de Fisica, Universidad Católica del Norte, Av. Angamos 0610, Antofagasta, Chile.

[3] Depto de Ciencias Farmaceúticas, Facultad de Ciencias, Universidad Católica del Norte, Antofagasta, Chile.

[4] Center for the Development of Nanoscience and Nanotechnology, CEDENNA, Santiago, Chile.

[5] Quantum Condensed Mater Division, Oak Ridge National Laboratory, 37830, Oak Ridge TN, USA

[6] Graduate School Materials Science in Mainz, Staudingerweg 9, 55128 Mainz, Germany

[*]Corresponding author: cemigdio@uni-mainz.de

**Abstract**. A thorough understanding of the mean-free-path (MFP) distribution of the energy carriers is crucial to engineer and tune the transport properties of materials. In this context, a significant body of work has investigated the phonon and electron MFP distribution, however, similar studies of the magnon MFP distribution have not been carried out so far. In this work, we used thickness-dependence measurements of the longitudinal spin Seebeck (LSSE) effect of yttrium iron garnet films to reconstruct the cumulative distribution of a SSE related effective magnon MFP. By using the experimental data reported by Guo et al. [Phys. Rev. X **6**, 031012 (2016)], we adapted the phonon MFP reconstruction algorithm proposed by A.J. Minnich, [Phys. Rev. Lett. **109**, 205901 (2012)] and apply it to magnons. The reconstruction showed that magnons with different MFP contribute in different manner to the total LSSE and the effective magnon MFP distribution spreads far beyond their typical averaged values.

**Introduction**

The continuous trend towards miniaturization of electronic components has allowed for more processing power at smaller dimensions. Consequently, large energy losses associated with the miniaturization have increased and cooling strategies have had to be developed. For this, thermoelectric devices have emerged as a promising green strategy for both cooling and energy harvesting. However, real applications are still limited by the low device efficiency and/or the toxicity of the component materials [1].

The quest for a new degree of freedom to improve the thermoelectric performance has opened up a new research area that includes the spin electron-heat interaction or spin caloritronics. Analogous to the conventional Seebeck mechanism, the spin Seebeck effect (SSE) describes the generation of spin currents as a result of a thermal gradient ($\nabla T$) across the material and it has been observed in metals, semiconductors and even in magnetic insulators [2]. In the longitudinal (or cross-plane) SSE configuration the generated spin currents are injected into a normal metal (NM) top layer and are electrically detected by measuring the inverse spin Hall effect [3–5]. This resultant voltage can be explained in terms of a thermal spin pumping caused by the temperature difference between the magnons in the ferromagnet (FM) and the electrons in the NM [6].

Several reports have demonstrated a very long effective (or averaged) magnon mean-free-path (M-MFP) in the magnetic insulator yttrium iron garnet (YIG) [7–12]. Typically, the estimation of the M-MFP is extracted from the best fit of (*i*) the thickness-dependence of the local SSE [9,10] or (*ii*) the separation-dependence of the injector-detector utilizing the nonlocal SSE [8,11]. Despite the fact that the fitting approach shows a very good match between the models and the experiments, recent works have demonstrated that the use of a single-averaged MFP may be inadequate and carriers with different mean free paths have different contributions to the total transport properties [13–20].

Recently, Cuffe et al. [14] showed that by using the thickness-dependence of the thermal conductivity of silicon membranes it is possible to reconstruct the phonon-MFP dependence of the Si bulk thermal conductivity. Based on the work of Minnich [15] and Yang and Dames [19], they formulated an inverse problem of reconstructing the phonon-MFP distribution from the experimental measurements by using the known relation between the membrane thickness and phonon-MFP suppression. Their results not only showed very good agreement with other more complicated theoretical approaches (e.g., molecular dynamics and *ab-initio* calculations), but also, and more importantly, they demonstrated that it is possible to reconstruct the MFP distribution from the experimental data by knowing the characteristic suppression function of the system.

Inspired by this work, we present an extension of the phonon-MFP reconstruction algorithm to reconstruct the M-MFP from SSE measurements (SSE-M-MFP) and spin diffusion length distribution (SDL). By using the thickness-dependence of the longitudinal SSE (LSSE) on YIG films measured by Guo et al. [10] and different suppression functions, we demonstrate that the SSE-M-MFP shows a broad distribution in YIG.

**Results and discussion**

In a bulk system, it is well-known that the different scattering events introduced by particles and quasi-particles, impurities, defects and dislocations, to name a few contributions, are the main impediment for the different carriers that determinate the transport properties in a material. However, reducing the size of the system, the surface scattering mechanism is one additional term that has a strong influence on the transport and which alters the physical properties as compared with the bulk. In general, the modelling of the modification of the transport properties due to the finite size effect is typically approached by adding the carrier-boundary scattering term ($\tau_B$) to the total relaxation time of the carrier ($\tau_{total}$), i.e., $1/\tau_{total} = 1/\tau + 1/\tau_B$., where $\tau$ is the lifetime of the carrier which contains all the intrinsic and extrinsic scattering mechanisms. However, Fuchs [21] showed that the boundary scattering is a surface phenomenon and the addition of an extra term in the total relaxation time is not strictly rigorous. Instead, it is necessary to include the boundary effect on the carrier mean free path. By solving the electronic-Boltzmann transport equation (BTE), Fuchs derived the effective electronic conductivity of a thin film subject to partially diffuse boundaries. This effect led to modification of the electronic distribution function, resulting in a modified formulation of the electronic conductivity. Years later, Chambers extended the model to nanowires [22] and Sondheimer simplified it to thin film and nanowires [23].

While this model was originally developed to describe the in-plane electrical conductivity of metallic thin-films, it has been extended to calculate: thermal conductivity [14,24,25], Hall, Seebeck and Peltier coefficients [26], the skin effect [27] and anisotropic magnetoresistance [28], among other properties. The Fuchs-Sondheimer (FS) model assumes that a fraction $p$ of carriers is specularly reflected and $1 - p$ carriers are diffusively reflected by the surface, respectively. Considering pure diffusive scattering, i.e., $p = 0$, the transport property of the thin-film, $\alpha_{film}$, can be related to the transport property of the bulk, $\alpha_{bulk}$, by:

$$\alpha_{film} = \alpha_{bulk} S(\chi) \qquad (1)$$

where $S(\chi)$ is the FS suppression function, $\chi = \Lambda/d$ is the Knudsen number, $\Lambda$ is the bulk-MFP and $d$ the film thickness. For the in-plane transport configuration the FS suppression function is given by:

$$S(\chi) = 1 - \frac{3}{8}\chi + \frac{3}{2}\chi \int_1^\infty (y^{-3} - y^{-5}) e^{-y/\chi} dy \qquad (2)$$

while for the cross-plane geometry it is expressed by [29]:

$$S(\chi) = 1 - 3\chi \left( \frac{1}{4} - \int_0^1 y^3 e^{-1/(\chi y)} dy \right) \qquad (3)$$

The Knudsen number-dependence of the cross-plane FS suppression function is displayed in Figure 1a.

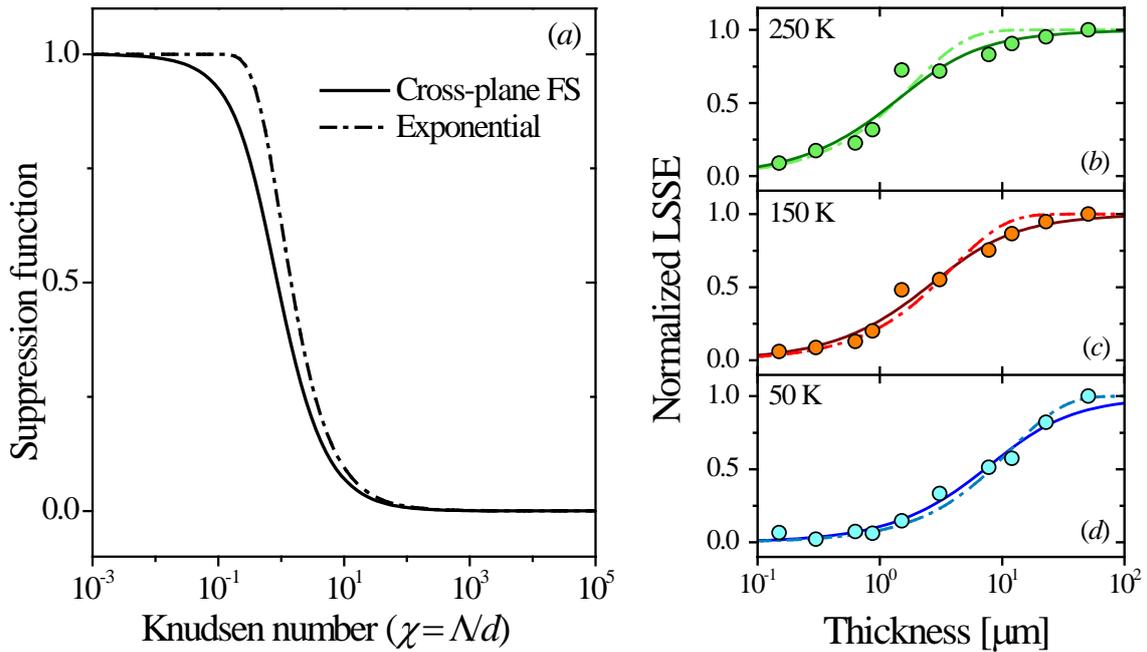

**Figure 1** (a) Knudsen number ($\chi$) dependence of cross-plane Fuchs-Sondheimer and exponential-like suppression function displayed in solid- and dashed-black lines, respectively.
(b-d) Experimental longitudinal spin Seebeck effect (LSSE) as a function of the YIG thickness measured at temperatures T = 250 (b), 150 (c) and 100 (d) K (green, orange and cyan dots, respectively), normalized to the thickest film sample. The experimental data were fitted by using cross-plane Fuchs-Sondheimer (solid lines) and exponential-like (dashed line) suppression functions.

For one-dimensional magnon transport subject to temperature gradient, Ritzmann et al. [30] showed that, by using an atomistic spin model with the stochastic Landau-Lifshitz-Gilbert equation as the underlying equation of motion, the spatial magnon accumulation ($\Delta m(z)$) decays exponentially on a length scale as $\Delta m(z) \sim \exp(-z/\Lambda)$. Later, Kehlberger et al. showed that the thickness dependence (suppression function) of the LSSE follows a similar trend given by [9]:

$$\frac{\alpha_{film}}{\alpha_{bulk}} = S(\chi) = 1 - \exp(-1/\chi) \qquad (4)$$

Typically, the best fitting curves to the experimental data are used to extract an averaged MFP. As an example, the best fitting curves of the LSSE in YIG films are shown in the Figure 1 (b-d). The experimental data was obtained from Ref [10] and they were normalized using the value of the thickest film sample $d = 55$ μm. It is important to mention that the exponential fitting was carried out by adjusting just the averaged MFP, which is different from Ref [10] where the amplitude of the fitting was also varied, leading to small differences in the fitted values. A summary of the averaged SSE-M-MFP, $\Lambda_{av}$, for different suppression functions and at different temperatures is displayed in Table 1. It is important to remark that cross-plane FS-like suppression function (Eq. 3) was calculated for phonon transport [29]. As the magnons can be treated as quasiparticles too, we considered that the BTE can be directly applied to the magnon transport and their modifications due to the introduction of boundaries. This function can be considered as an extension to the one-dimensional magnon transport (Eq. (4)), because FS-like suppression function integrates over the full space taking into account the vector properties of individual carriers. For a detailed description on the derivation of the in- and cross-plane FS suppression functions, the readers are referred to the work of Cuffe et al. [14] and Hua and Minnich [29], respectively.

**Table 1**. Averaged SSE related magnon mean free path ($\Lambda_{av}$) for different temperatures and suppression functions.

| T [K] | Cross-plane FS (Eq. 3) | Exponential-like (Eq. 4) |
|---|---|---|
|  | $\Lambda_{av}$ [μm] | $\Lambda_{av}$ [μm] |
| 250 | 1.1 | 1.9 |
| 200 | 1.6 | 2.8 |
| 150 | 2.1 | 3.9 |
| 100 | 3.1 | 5.7 |
| 50 | 6.3 | 11.7 |

Despite the good match between the different models and the experiment, recent works have demonstrated that the use of a single averaged MFP may be inadequate and carriers with different mean free paths have different contributions to the total transport properties [13–16].

Taking into account this effect, Minnich [15] suggested a method to extract the MFP accumulation function based on a characteristic suppression function, which depends on the geometry and the experimental configuration. For the particular case of the thermal conductivity of a thin Si membrane, Cuffe et al. [14] showed that the measured thermal conductivity of a film ($\alpha_{film}(d)$) and the suppression function are related through a cumulative MFP distribution given by [14,15]:

$$\alpha_{film}(d) = \int_0^\infty S(\chi) f(\Lambda) d\Lambda \tag{5}$$

$$\alpha = \frac{\alpha_{film}(d)}{\alpha_{bulk}} = \int_0^\infty K(\chi) F_{acc}(\Lambda) d\Lambda, \quad \text{with} \tag{6}$$

$$F_{acc}(\Lambda c) = \frac{1}{\alpha_{bulk}} \int_0^{\Lambda c} f(\Lambda') d\Lambda', \quad \text{and} \quad K(\chi) = -\frac{dS(\chi)}{d\chi} \frac{d\chi}{d\Lambda} \tag{7}$$

The $F_{acc}(\Lambda c)$ represents the fraction of transport property contribution from all the carriers with MFP less than $\Lambda c$ (in our case the cumulative bulk LSSE) and $K(\chi)$ represents the computational Kernel of the integral. Then, to recover $F_{acc}(\Lambda)$ from the experimental data it is necessary to have a wide enough range of measurements at different film thicknesses. The solution of Eq. (6) is technically an ill-posed problem and, in principle, with infinite solutions. However, Minnich [15] showed that one can discretize the integral and impose some constrains to the $F_{acc}$ to obtain a unique solution. These conditions are mainly related to the "shape" of the reconstructed function and they are expressed in terms of a minimization problem through the Tikhonov regularization method given by:

$$\min \left\{ \|A \cdot F - \alpha\|_2^2 + \mu^2 \|L \cdot F\|_2^2 \right\} \tag{8}$$

where $\| \|_2$ is the second-norm operator, $\alpha_i$ is the normalized LSSE of the $i^{th}$ measurement, $A_{i,j} = K(\chi_{i,j}) \cdot \beta_j$ is a $m \times n$ matrix with $m$ the number of measurements and $n$ the discretization points of the integral, $\beta_j$ is the weight of the quadrature point at $\chi_{i,j} = \Lambda_j / d_i$, $F$ is a vector of the desired accumulation function ($F_{acc}$), $\mu$ is the regularization parameter to control the smoothness of $F$ and $L$ is a $(n-2) \times n$ tridiagonal Toeplitz matrix which represents an approximation of a second derivative operator, i.e., $L \cdot F = F_{i+1} - 2F_i + F_{i-1}$. Despite that the context of our problem is completely different to the proposed by Minnich, these constrains are directly transferable to this reconstruction. Due to the magnons dispersion relation, group velocity and density of states do not show abrupt singularities [31,32], therefore, we also expect their MFP-distribution will be a smooth function without abrupt jumps which increases monotonically. In addition, we reconstruct the integrated distribution, which will be comparably smooth in spite of possible singularities in the actual distribution.

Finally, to numerically solve the bulk effective SSE-M-MFP reconstruction, we have discretized the integral of Eq. (6) through a trapezoid method using 50 logarithmic-spaced points with a regularization parameter $\mu = 0.85$ and 1.2 for the suppression functions defined in Eq. (3) and (4), respectively, which describe the cross-plane (longitudinal) SSE transport. The minimization of Eq. (8) was performed by using CVX, a package for specifying and solving convex programs [33,34], and the experimental data of the LSSE in YIG films reported by Guo et al. [10]. The regularization parameters were estimated through *L-curve criterion* [35,36] using the plot of the norm of the regularised solution ($\|L \cdot F\|_2$) versus the norm of the residual norm ($\|A \cdot F - b\|_2$). A brief description of this method and one example of the *L-curve* for FS suppression function can be found in the supplementary information.

The reconstructed bulk M-MFP for three different temperatures is shown in Figure 2. The M-MPF distribution was calculated using a cross-plane Fuchs-Sondheimer and an exponential-like suppression function represented by solid and dashed lines, respectively. The solid-circles and solid-squares show the MFP of magnons which contribute 50 % to the overall LSSE in YIG.

We notice that depending on the selected suppression function the effective SSE-M-MFP spans different ranges. This difference is basically due to the mathematical nature of each suppression function used in the reconstruction. Due to that the MFP reconstruction is an ill-posed problem a drastic change of the suppression function can have a strong impact on their MFP distribution. For the FS-like suppression function (Eq. (3)) the SSE-M-MFP ranges from ~ 0.4-4.5 µm, 0.9-9 µm and 1.6-20.3 µm for T = 250, 100 and 50 K, respectively, contributing between 10-90% to the overall LSSE. Whereas for the exponential-like function (Eq. (4)) the SSE-M-MFP shows a broader distribution spanning from ~ 0.7-11 µm, 1.9-17 µm and 3-34 µm for the same temperature values and percentages of contribution.

For both the FS and the exponential suppression function, we observe that the SSE-M-MFP does not show a single value and it is extended far beyond to their averaged value (see Table 1). In a similar way than phonons, MFPs longer than 1.3-2.2 µm (see the solid-circles and solid-squares in the Figure

2) contribute 50% to the total LSSE, and, as we go down in temperature, magnons with longer MFPs become relatively more important and the SSE-M-MFP distribution becomes broader.

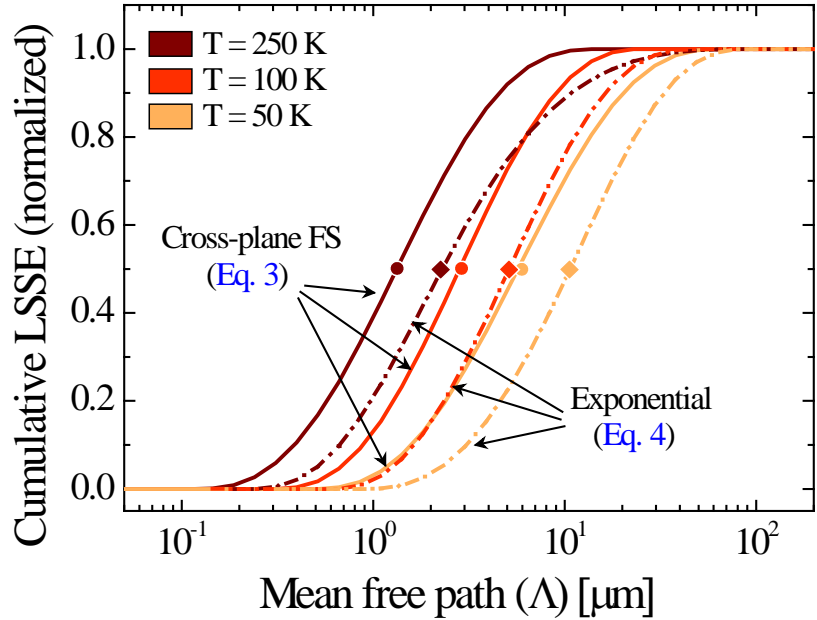

**Figure 2** Reconstructed magnon MFP distribution for YIG using cross-plane Fuchs-Sondheimer (FS, solid lines) and exponential (dashed lines) reduction functions at three different temperatures T = 250, 100 and 50 K, respectively. The solid-circles and solid-squares represent the SSE-M-MFP carrying 50% of the LSSE for FS and exponential reduction function respectively.

The large distribution of the SSE-M-MFP will have a strong impact on the maximum value of the LSSE at low temperatures. As it was reported by Guo et al. [10], the maximum LSSE peak shows a strong dependence on the film thicknesses. It shifts to higher temperatures as the thickness decreases, disappearing completely for thickness below 1 μm. This effect indicates that magnons dissipate their energy mainly by boundary scattering, and, as a consequence, in a similar manner to phonon-transport, the boundary will impose an upper limit to the MFP. Then, the maximum of the LSSE will shift to higher temperatures, i.e., magnons with a shorter MFP, until it disappears completely in the ultra-thin film limit [37].

It is important to remark that the effective SSE-M-MFP values derived from the LSSE will be valid for this system of Pt (5 nm)/YIG. The LSSE is also dependent on the interface quality of NM/YIG [9,10,38] and the thickness of NM layer [39–41], which has its own spectral-distribution on the NM. Taking into account this effect we also have calculated the spectral distribution of the spin diffusion length (SDL) of Pt. Based on the thickness dependence of the spin Hall torque efficient, $\xi$, reported by Nguyen et al. [41] for Co/Pt bilayers and the functional suppression function given by [42]:

$$\frac{\xi(d)}{\xi_{max}} = 1 - \frac{1}{\sinh(1/\chi)}, \qquad (9)$$

the SDL distribution of Pt is presented on Figure 3.

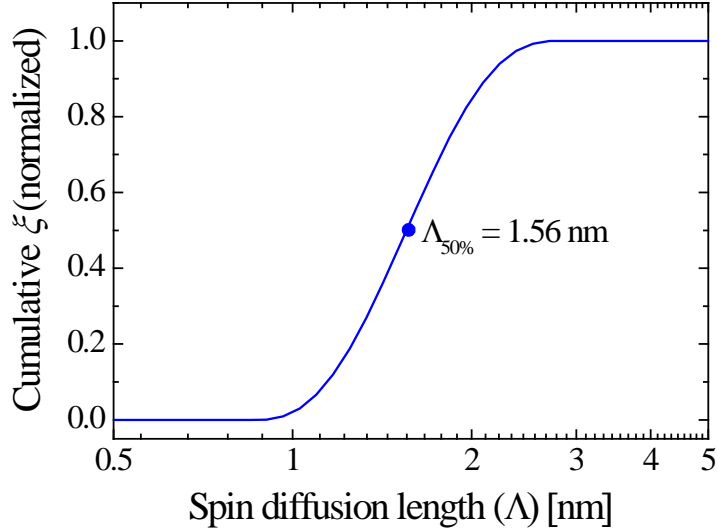

**Figure 3** Room-temperature reconstructed spin diffusion length of Pt films. The reconstruction was carried out using Eq. (9) and the thickness dependence of the spin Hall torque coefficient reported in Ref [41].

The regularization parameter was also estimated through *L-curve criterion* with an optimal value of $\mu = 0.78$.

In contrast to the effective SSE-M-MFP distribution, the SDL distribution for Pt spans in a narrower region between 1.3-2 nm for 10 to 90% contribution. It indicates that the use of a single averaged value of the spin diffusion length is indeed a good approximation, and that the Pt thickness in the detector stripes of Guo et al. (5 nm) is thick enough to get the full SSE signal. However, it is important to mention that Eq. (9) assumes that the resistivity and spin-Hall angle of the non-magnetic metal layer are independent of the thickness. Therefore, this is a first approximation and the resistivity and spin-Hall angle thickness dependence will also play a role in the form of the suppression function and in the distribution of the spin diffusion length. In this context, a deeper study of this function is needed to obtain a complete insight into the distribution of the SDL.

**Conclusions**

Using the experimental size-dependence of the spin Seebeck effect in YIG films and the phonon MFP-reconstruction algorithm, we showed that by an extension of this concept applied to the LSSE data, one can calculate SSE related magnon MFP distributions. The reconstruction was performed using cross-plane Fuchs-Sondheimer and exponential suppression functions. Both functions showed that the effective SSE-M-MFP distribution spreads a range of values (at least two order of magnitude) far beyond the averaged value. Note that this effective M-MFP derived from the SSE measurements will be valid for a given system consisting of Pt (5 nm)/YIG including the specific interface and thickness of Pt layer. Consequently one needs to take into account the spin-charge conversion processes at the interface and the results need to be considered as effective values valid for this system, however the numerical procedure can be applied to any configuration. By developing spectrally independent spin charge conversion of the spin wave transmission at the interface, combined with a spectrally independent inverse spin Hall effect this analysis can provide the intrinsic M-MFP.

Additionally, using the same algorithm the distribution of the spin diffusion length was also carried out. In contrast to magnons, the spin diffusion length showed a narrow distribution, indicating that the single averaged values can be used as a good approximation in this case and demonstrating that the NM thickness of the detector used for the effective M-MFP reconstruction will not affect the LSEE signal.

**Acknowledgments**

We gratefully acknowledge financial support by the Deutsche Forschungsgemeinschaft, DFG, Germany, [Grants No. Ja821/4 within SPP 1386 (Nanostructured Thermoelectric Materials) and No. Ja821/7-1 and KL1811/7-2 within SPP 1538 (Spin Caloric Transport)], the Transregional Collaborative Research Center SFB/TRR173 "Spin+X– Spin its collective environment" and the EU project InSpin [Grant No FP7-ICT-2013-612759]. R.Z.A. gratefully acknowledges financial support from ACT1204 [ANILLO DE INVESTIGACIÓN] and SF acknowledges financial support from CEDENNA FB0807. We thank Dr. John Cuffe and Dr. Ulrike Ritzmann for the valuable discussions. E.C.A expresses his gratitude to Ines Puertas for all the inspiring discussions during the development of this work.

# Supplementary information: Estimation of the regularization parameter

The estimation of the regularization parameter ($\mu$) was estimated using the L-curve criterion [S1-S3]. The L-curve criterion is a widely used tool for choosing the most adequate regularization parameter in an ill-posed (or inverse) problems of the form:

$$\min\left\{ \|A \cdot F - \alpha\|_2^2 + \mu^2 \|L \cdot F\|_2^2 \right\} \quad (S10)$$

where $A$ is the ill-conditioned matrix which, in our problem, represents the part of the discretised integral defined in Eq. (6), $F$ is the cumulative bulk LSEE, $\alpha_i$ is the normalized LSSE of the $i^{th}$ measurement, $\mu$ is the regularization parameter and $L$ is a second derivative operator, i.e., $L \cdot F = F_{i+1} - 2F_i + F_{i-1}$.

The first element of the Eq. S1 ensures the solution of the discretised integral defined in Eq. (6) and the second one enforces smoothness in the reconstructed distribution ($F_{acc}$) which is controlled by the regularization parameter $\mu$.

The selection of most adequate $\mu$ is a hot topic of research in mathematics and several heuristic methods have been applied for this task, such as, e.g.: discrepancy principle, generalized cross-validation (GCV), *L-curve* criterion, the Gfrerer/Raus-method, the quasi-optimality method, to name a few [4]. Among these methods, the *L-curve* criterion is one of the most popular one due to its robustness, velocity and efficiency. Basically, this method balances the size of the discrepancy in the solution (residual norm, $\|A \cdot F_\mu - \alpha\|_2$) with the size of the solution (solution norm, $\|L \cdot F_\mu\|_2$) for different values of $\mu$. As is displayed in Fig. S1b, the curve has an *L*-like shape composed by a flat and steep parts. The flat part represents solutions dominated by regularization errors and the steep perturbation errors, respectively. The corner represents a compromise between the data fitting and the smoothness of the solution [S1-S3]. Hansen showed that one fast method to obtain the corner point in the *L-curve* is through the analysis of its curvature [S2], which is given by:

$$k(\mu) = 2 \frac{\left| \rho'(\mu)\eta''(\mu) - \rho''(\mu)\eta'(\mu) \right|}{\left( \rho'(\mu)^2 + \eta'(\mu)^2 \right)^{3/2}} \quad (S11)$$

where $\eta(\mu) = \|L \cdot F_\mu\|_2$, $\rho(\mu) = \|A \cdot F_\mu - \alpha\|_2$ and $\eta'$, $\rho'$, $\eta''$ and $\rho''$ are the first and second derivative of $\eta$ and $\rho$ with respect to regularization parameter $\mu$, respectively. It is important to mention that in the original work of Hansen $\eta$ and $\rho$ are described in terms of the logarithm of the solution norm and residual norm, in concordance with *L-criterion*. However, in our case, we did not see very large variation of the solution norm, and the double logarithmic scaling does not yield improved results.

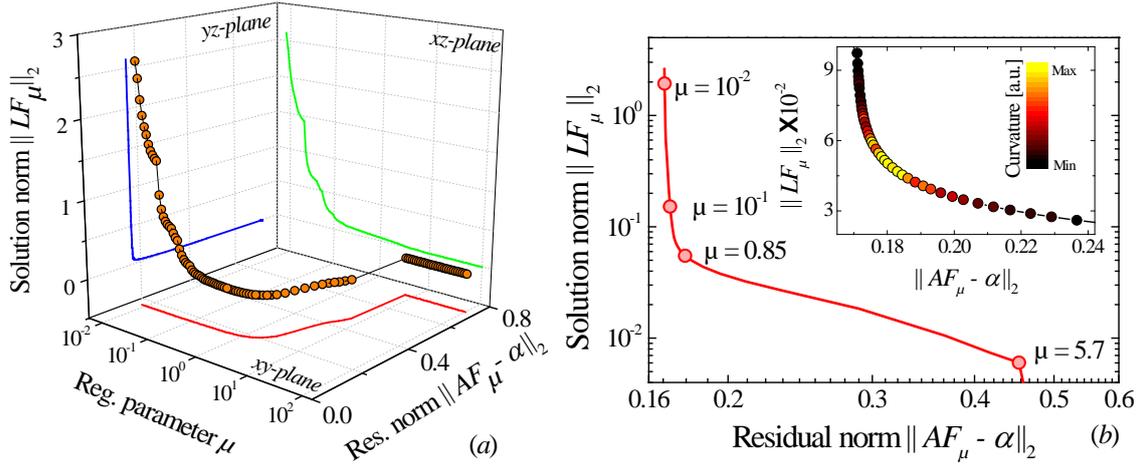

**Figure S4** (a) Three-dimensional like representation of the solution norm ($\|LF_\mu\|_2$) as a function of the regularization parameter ($\mu$) and the residual norm ($\|LF_\mu - \alpha\|_2$). (b) Computed L-curve (or *zy* projection of (a)), different regularization parameters are marked with red solid dots. In the inset the curvature values are displayed in heat-like colour bar.

Figure S1a shows the three-dimensional-like (3D) representation of solution norm as a function of the regularization parameter and the residual norm. The 3D-like curve was computed using Fuchs-Sondheimer suppression function given in Eq. 3 and the LSSE experimental data measured at 250K. Some interesting features from this graph are represented in the different projections in the *xy*, *xz* and *yz* planes. The *xy* and *xz* projections give the variation of the residual norm and solution norm for different regularization parameters. We can see that very small $\mu$-values introduce large errors in the solution norm, while large values of $\mu$ introduce large errors in the residual norm.

Finally, the *yz* projection is the computed *L-curve*. A detailed graph is showed in Figure S1b, where different values of the $\mu$-parameter are displayed with red-solid dots and the curvature is presented in the inset. The heat-like colour bar represents the maximum (yellow) and minimum (black) values of the curvature. We can observe that the maximum of the curvature is located in the corner of the *L-curve* for $\mu \approx 0.85$.

Once the $\mu$-parameter has been set, the reconstructed function is calculated as shown in Figure S2. The optimal $\mu$-value is displayed in red-solid line, while small deviation around 10% of the maximum curvature is shown as light-red-like shadow.

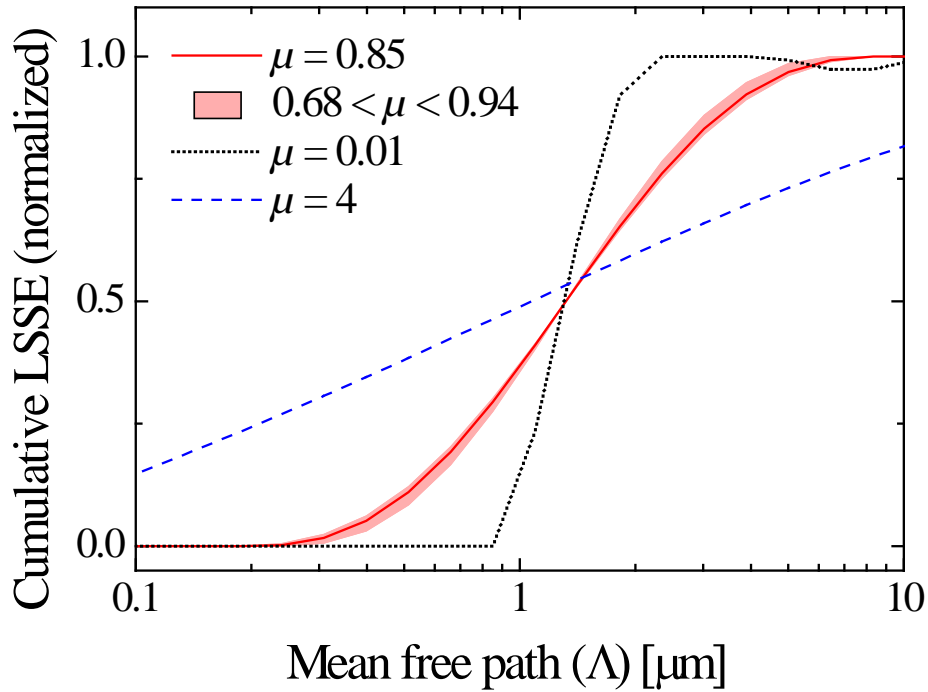

**Figure S5** Reconstructed magnon LSEE-related MFP distribution for YIG using cross-plane Fuchs-Sondheimer and the experimental LSSE data measured at 250K. The light-red-like shadow represents 10% variation of the optimal regularization parameter ($\mu = 0.85$, red solid line) in range of $0.68 < \mu < 0.94$. Other $\mu$-values are also plotted with black-dotted ($\mu = 0.08$) and blue-dashed ($\mu = 4$) lines as an example.

As we can see in the reconstructed distribution the small deviation from the optimal value do not change significant the reconstructed function. However, larger deviation from the optimum value (blue-dashed and black-dotted lines) can introduce corresponding large deviations in the reconstructed distribution.